%
%
\input harvmac %
\input epsf
%
%
%
%
%

%
%
%
%
\newif\ifdraft

\noblackbox
\catcode`\@=11
\newif\iffrontpage
%
\ifx\answ\bigans
\def\titleft{\titsm}
\magnification=1200\baselineskip=15pt plus 2pt minus 1pt
%
\advance\hoffset by-0.075truein
\hsize=6.15truein\vsize=600.truept\hsbody=\hsize\hstitle=\hsize
\else\let\lr=L
\def\titleft{\titla}
\magnification=1000\baselineskip=14pt plus 2pt minus 1pt
%
\vsize=6.5truein
\hstitle=8truein\hsbody=4.75truein
\fullhsize=10truein\hsize=\hsbody
\fi
\parskip=4pt plus 10pt minus 4pt

\font\titla=cmr10 scaled\magstep3
\font\tenmss=cmss10
\font\absmss=cmss10 scaled\magstep1
\newfam\mssfam
\font\footrm=cmr8  \font\footrms=cmr5
\font\footrmss=cmr5   \font\footi=cmmi8
\font\footis=cmmi5   \font\footiss=cmmi5
\font\footsy=cmsy8   \font\footsys=cmsy5
\font\footsyss=cmsy5   \font\footbf=cmbx8
\font\footmss=cmss8
\def\footfont{\def\rm{\fam0\footrm}
\textfont0=\footrm \scriptfont0=\footrms
\scriptscriptfont0=\footrmss
\textfont1=\footi \scriptfont1=\footis
\scriptscriptfont1=\footiss
\textfont2=\footsy \scriptfont2=\footsys
\scriptscriptfont2=\footsyss
\textfont\itfam=\footi \def\it{\fam\itfam\footi}
\textfont\mssfam=\footmss \def\mss{\fam\mssfam\footmss}
\textfont\bffam=\footbf \def\bf{\fam\bffam\footbf} \rm}
\def\tenpoint{\def\rm{\fam0\tenrm}
\textfont0=\tenrm \scriptfont0=\sevenrm
\scriptscriptfont0=\fiverm
\textfont1=\teni  \scriptfont1=\seveni
\scriptscriptfont1=\fivei
\textfont2=\tensy \scriptfont2=\sevensy
\scriptscriptfont2=\fivesy
\textfont\itfam=\tenit \def\it{\fam\itfam\tenit}
\textfont\mssfam=\tenmss \def\mss{\fam\mssfam\tenmss}
\textfont\bffam=\tenbf \def\bf{\fam\bffam\tenbf} \rm}
\ifx\answ\bigans\def\abstractfont{\tenpoint}\else
\def\abstractfont{\def\rm{\fam0\absrm}
\textfont0=\absrm \scriptfont0=\absrms
\scriptscriptfont0=\absrmss
\textfont1=\absi \scriptfont1=\absis
\scriptscriptfont1=\absiss
\textfont2=\abssy \scriptfont2=\abssys
\scriptscriptfont2=\abssyss
\textfont\itfam=\bigit \def\it{\fam\itfam\bigit}
\textfont\mssfam=\absmss \def\mss{\fam\mssfam\absmss}
\textfont\bffam=\absbf \def\bf{\fam\bffam\absbf}\rm}\fi
%
\def\f@@t{\baselineskip10pt\lineskip0pt\lineskiplimit0pt
\bgroup\aftergroup\@foot\let\next}
\setbox\strutbox=\hbox{\vrule height 8.pt depth 3.5pt width\z@}
\def\vfootnote#1{\insert\footins\bgroup
\baselineskip10pt\footfont
\interlinepenalty=\interfootnotelinepenalty
\floatingpenalty=20000
\splittopskip=\ht\strutbox \boxmaxdepth=\dp\strutbox
\leftskip=24pt \rightskip=\z@skip
\parindent=12pt \parfillskip=0pt plus 1fil
\spaceskip=\z@skip \xspaceskip=\z@skip
\Textindent{$#1$}\footstrut\futurelet\next\fo@t}
\def\Textindent#1{\noindent\llap{#1\enspace}\ignorespaces}
\def\footnote#1{\attach{#1}\vfootnote{#1}}%

\def\foot{\attach\footsymbolgen\vfootnote{\footsymbol}}
\let\footsymbol=\star
\newcount\lastf@@t           \lastf@@t=-1
\newcount\footsymbolcount    \footsymbolcount=0
\def\footsymbolgen{\relax\footsym
\global\lastf@@t=\pageno\footsymbol}
\def\footsym{\ifnum\footsymbolcount<0
\global\footsymbolcount=0\fi
{\iffrontpage \else \advance\lastf@@t by 1 \fi
\ifnum\lastf@@t<\pageno \global\footsymbolcount=0
\else \global\advance\footsymbolcount by 1 \fi }
\ifcase\footsymbolcount \fd@f\star\or
\fd@f\dagger\or \fd@f\ast\or
\fd@f\ddagger\or \fd@f\natural\or
\fd@f\diamond\or \fd@f\bullet\or
\fd@f\nabla\else \fd@f\dagger
\global\footsymbolcount=0 \fi }
\def\fd@f#1{\xdef\footsymbol{#1}}
\def\space@ver#1{\let\@sf=\empty \ifmmode #1\else \ifhmode
\edef\@sf{\spacefactor=\the\spacefactor}
\unskip${}#1$\relax\fi\fi}
 \def\attach#1{\space@ver{\strut^{\mkern 2mu #1}}\@sf}
%
\newif\ifnref
\def\rrr#1#2{\relax\ifnref\nref#1{#2}\else\ref#1{#2}\fi}
\def\ldf#1#2{\begingroup\obeylines
\gdef#1{\rrr{#1}{#2}}\endgroup\unskip}

\nreffalse
\def\refout{\listrefs}
%
\def\eqn#1{\xdef #1{(\secsym\the\meqno)}
\writedef{#1\leftbracket#1}%
\global\advance\meqno by1\eqno#1\eqlabeL#1}
\def\eqnalign#1{\xdef #1{(\secsym\the\meqno)}
\writedef{#1\leftbracket#1}%
\global\advance\meqno by1#1\eqlabeL{#1}}
%
\def\chap#1{\newsec{#1}}
\def\chapter#1{\chap{#1}}
\def\sect#1{\subsec{{ #1}}}
\def\section#1{\sect{#1}}
\def\\{\ifnum\lastpenalty=-10000\relax
\else\hfil\penalty-10000\fi\ignorespaces}
\def\note#1{\leavevmode%
\edef\@@marginsf{\spacefactor=\the\spacefactor\relax}%
\ifdraft\strut\vadjust{%
\hbox to0pt{\hskip\hsize%
\ifx\answ\bigans\hskip.1in\else\hskip-.1in\fi%
\vbox to0pt{\vskip-\dp
\strutbox\sevenbf\baselineskip=8pt plus 1pt minus 1pt%
\ifx\answ\bigans\hsize=.7in\else\hsize=.35in\fi%
\tolerance=5000 \hbadness=5000%
\leftskip=0pt \rightskip=0pt \everypar={}%
\raggedright\parskip=0pt \parindent=0pt%
\vskip-\ht\strutbox\noindent\strut#1\par%
\vss}\hss}}\fi\@@marginsf\kern-.01cm}
\def\titlepage{%
\frontpagetrue\nopagenumbers\abstractfont%
\hsize=\hstitle\rightline{\vbox{\baselineskip=10pt%
{\abstractfont\pubnum}}}\pageno=0}
\frontpagefalse
\def\pubnum{}
\def\pdate{\number\month/\number\yearltd}
\def\makefootline{\iffrontpage\vskip .27truein
\line{\the\footline}
\vskip -.1truein\leftline{\vbox{\baselineskip=10pt%
{\abstractfont\pdate}}}
\else\vskip.5cm\line{\hss \tenrm $-$ \folio\ $-$ \hss}\fi}
\def\title#1{\vskip .7truecm\titlestyle{\titleft #1}}
\def\titlestyle#1{\par\begingroup \interlinepenalty=9999
\leftskip=0.02\hsize plus 0.23\hsize minus 0.02\hsize
\rightskip=\leftskip \parfillskip=0pt
\hyphenpenalty=9000 \exhyphenpenalty=9000
\tolerance=9999 \pretolerance=9000
\spaceskip=0.333em \xspaceskip=0.5em
\noindent #1\par\endgroup }
\def\autskip{\ifx\answ\bigans\vskip.5truecm\else\vskip.1cm\fi}
\def\author#1{\vskip .7in \centerline{#1}}

\def\address#1{\ifx\answ\bigans\vskip.2truecm
\else\vskip.1cm\fi{\it \centerline{#1}}}
\def\abstract#1{
\vskip .3in\vfil\centerline
{\bf Abstract}\penalty1000
{{\smallskip\ifx\answ\bigans\leftskip 2pc \rightskip 2pc
\else\leftskip 5pc \rightskip 5pc\fi
\noindent\abstractfont \baselineskip=12pt
{#1} \smallskip}}
\penalty-1000}
\def\endpage{\tenpoint\supereject\global\hsize=\hsbody%
\frontpagefalse\footline={\hss\tenrm\folio\hss}}
%


\def\bfone{\relax{\rm 1\kern-.35em 1}}
\def\inbar{\vrule height1.5ex width.4pt depth0pt}
\def\IC{\relax\,\hbox{$\inbar\kern-.3em{\mss C}$}}
\def\ID{\relax{\rm I\kern-.18em D}}
\def\IF{\relax{\rm I\kern-.18em F}}
\def\IH{\relax{\rm I\kern-.18em H}}
\def\II{\relax{\rm I\kern-.17em I}}
\def\IN{\relax{\rm I\kern-.18em N}}
\def\IP{\relax{\rm I\kern-.18em P}}
\def\IQ{\relax\,\hbox{$\inbar\kern-.3em{\rm Q}$}}
\def\IR{\relax{\rm I\kern-.18em R}}
\font\cmss=cmss10 \font\cmsss=cmss10 at 7pt
\def\ZZ{\relax\ifmmode\mathchoice
{\hbox{\cmss Z\kern-.4em Z}}{\hbox{\cmss Z\kern-.4em Z}}
{\lower.9pt\hbox{\cmsss Z\kern-.4em Z}}
{\lower1.2pt\hbox{\cmsss Z\kern-.4em Z}}\else{\cmss Z\kern-.4em
Z}\fi}

\def\t1{t_1}
\def\t2{t_2}
\def\t3{t_3}
\def\t4{t_4}
\def\t5{t_5}
\def\nup#1({Nucl.\ Phys.\ $\us {B#1}$\ (}
\def\plt#1({Phys.\ Lett.\ $\us  {#1}$\ (}
\def\cmp#1({Comm.\ Math.\ Phys.\ $\us  {#1}$\ (}
\def\prp#1({Phys.\ Rep.\ $\us  {#1}$\ (}
\def\prl#1({Phys.\ Rev.\ Lett.\ $\us  {#1}$\ (}
\def\prv#1({Phys.\ Rev.\ $\us  {#1}$\ (}
\def\mpl#1({Mod.\ Phys.\ \Let.\ $\us  {#1}$\ (}
\def\tit#1|{{\it #1},\ }
%

%

\def\us#1{\underline{#1}}

\def\Coe#1.#2.{{#1\over #2}}
\def\coeff#1#2{\relax{\textstyle {#1 \over #2}}\displaystyle}
\def\coe#1.#2.{\relax{\textstyle {#1 \over #2}}\displaystyle}

\def\shalf{\relax{\textstyle {1 \over 2}}\displaystyle}

\def\to{\rightarrow}
\def\notin{\hbox{{$\in$}\kern-.51em\hbox{/}}}

\catcode`\@=12
%
\def\note#1{\ifdraft {\bf [#1]} \else\fi}
\def\cref{\ifdraft {\bf [check refs]}\else\fi}

\def\undertext#1{\vtop{\hbox{#1}\kern 1pt \hrule}}
\def\subsection#1{\vskip 0.5cm\undertext{\rm #1}}

\def\ap{\alpha'}
\def\aph{\alpha_{\rm h}'}

\def\Int#1#2{{\textstyle\int_{#1}^{#2}}}

\parindent0pt
%
\def\hepth#1{{\tt hep-th}/#1}

\ldf\sagnotti{A.Sagnotti, Phys. Lett. B294 (1992) 196}

\ldf\wvd{E. Witten, ``String Theory Dynamics in Various Dimensions",
Nucl. Phys. B443 (1995) 85, \hepth9503124}

\ldf\pw{J. Polchinski and E. Witten, ``Evidence for Heterotic-Type I
String Duality'',  
Nucl. Phys. B460 (1996) 525, \hepth9710169}

\ldf\leigh{J. Dai, R.G. Leigh and J. Polchinski, 
``New Connections Between String Theories",
Mod. Phys. Lett. A, Vol. 4, 21 (1989) 2073}

\ldf\ginsp{P. Ginsparg, ``Comment on Toroidal Compactification of
Heterotic Superstrings",
Phys. Rev. D35 (1986) 648}

\ldf\hwone{P. Horava and E. Witten, 
``Heterotic and Type I String Dynamics from Eleven Dimensions",
Nucl. Phys. B460 (1996) 506, \hepth9510209}

\ldf\dafe{U. Danielsson and G. Ferretti, 
``The Heterotic Life of the D Particle", 
Int. J. Mod. Phys. A12 (1997) 4581, 
\hepth9610082}

\ldf\bfss{T. Banks, W. Fischler, S.H. Shenker and L. Susskind,
``M Theory as a Matrix Model: a Conjecture'', 
Phys. Rev. D55 (1997) 5112, \hepth9610043}

\ldf\motl{L. Motl, "Proposals on Non-Perturbative Superstring Interactions", 
hep-th/9701025}

\ldf\kire{N. Kim and S.J Rey, "M(atrix) Theory On An Orbifold And Twisted
Membrane", \hepth9701139}

\ldf\lowone{D. Lowe, 
"Bound States of type $I^\prime$ D--Particles And Enhanced
Gauge Symmetry" \hepth9702006}

\ldf\bss{T. Banks, N. Seiberg and E. Silverstein, "Zero and One--Dimensional
Probes With N=8 Supersymmetry", \hepth9703052}

\ldf\bamo{T. Banks and L. Motl, "Heterotic Strings From Matrices", 
\hepth9703218}

\ldf\lowtwo{D. Lowe, "Heterotic Matrix String Theory", \hepth9704041}

\ldf\rey{S.J Rey, "Heterotic M(atrix) Strings And Their Interactions",
\hepth9704158}

\ldf\horav{P. Horava, "Matrix Theory And Heterotic Strings On Tori",
\hepth9705055}

\ldf\pl{J. Polchinski, ``TASI Lectures on D--Branes", \hepth9611050}

\ldf\gene{E. Bergshoeff, M. de Roo, M.B. Green, G. Papadopoulos 
and P.K. Townsend, Nucl. Phys. B470 (1996) 113}

\ldf\wcy{E. Witten, 
``Strong Coupling Expansion of Calabi-Yau Compactifications'',
Nucl. Phys. B471 (1996) 135, \hepth9602070}

\ldf\bgl{O. Bergman, M.R. Gaberdiel and G. Lifschytz, 
``Branes, Orientifolds And The Creation Of Elementary Strings",
\hepth9705130}

\ldf\pcj{J. Polchinski, S. Chaudhuri and C. Johnson, 
``Notes on D-Branes", \hepth9602052}

\ldf\hetmatrix{
U. Danielsson and G. Ferretti, 
``The Heterotic Life of the D Particle", 
Int. J. Mod. Phys. A12 (1997) 4581, \hepth9610082;\hfill\break
S. Kachru and E. Silverstein, 
``On Gauge Bosons in the Matrix Model Approach to M Theory'', 
Phys. Lett. 396B (1997) 70, \hepth9612162; \hfill\break
T. Banks, N. Seiberg and E. Silverstein, ``Zero and One--Dimensional
Probes With N=8 Supersymmetry", 
Phys. Lett. B401 (1997) 30, \hepth9703052;\hfill\break
T. Banks and L. Motl, ``Heterotic Strings from Matrices", 
\hepth9703218;\hfill\break
D. Lowe, ``Bound States of Type I$'$ D Particles and Enhanced
Gauge Symmetry", Nucl. Phys. B501 (1997) 134, \hepth9702006;
``Heterotic Matrix String Theory", 
Phys. Lett. B403 (1997) 243, \hepth9704041;\hfill\break
S.J Rey, ``Heterotic M(atrix) Strings and their Interactions",
\hepth9704158;\hfill\break
P. Horava, ``Matrix Theory and Heterotic Strings on Tori",
\hepth9705055;\hfill\break
D. Kabat and S.-J. Rey, ``Wilson Lines and T-Duality in Heterotic
Matrix Theory", \hepth9707099}

\ldf\ks{S. Kachru and E. Silverstein, 
as in \hetmatrix}

\ldf\bd{J. Blum and K. Dienes,
``From the  Type I String to M Theory: a Continuous Connection",
\hepth9708016}  

\ldf\bk{C. Bachas and E. Kiritsis, ``$F^4$ Terms in N=4 String Vacua", 
Nucl. Phys. Proc. Suppl. 55B (1997), \hepth9611205}

\ldf\bfkov{C. Bachas, C. Fabre, E. Kiritsis, N.A. Obers and P. Vanhove, 
``Heterotic/Type I Duality and D-brane Instantons", \hepth9707126}

%
%
\footline{\hss\tenrm--\folio\--\hss}
\def\pubnum{\hbox{LMU-TPW-97-26}
\hbox{NEIP-97-012}}
\titlepage
\vskip 1cm
\title{\hfill Matching the BPS Spectra of \hfill\break
\vskip-.4cm
Heterotic - Type I - Type I$^{'}$ 
Strings$^{\diamond}$}
\author{Dimitris Matalliotakis$^{1,2}$, Hans-Peter Nilles$^{2,3,*}$, 
Stefan Theisen$^4$}
\vskip .7cm
\centerline{$^1$\it Institut de Physique Theorique}
\centerline{\it Universit\'e de Neuch\^atel, CH-2000 Neuch\^atel}
\medskip
\centerline{$^2$\it Institut f\"ur Theoretische Physik}
\centerline{\it Technische Universit\"at M\"unchen, D-85747 M\"unchen, FRG}
\medskip
\centerline{$^3$\it Max-Planck-Institut f\"ur Physik}
\centerline{\it D-80805 M\"unchen, FRG}
\medskip
\centerline{$^4$\it Sektion Physik, Universit\"at M\"unchen, FRG}
\medskip
\centerline{\footfont
dmatalli@physik.tu-muenchen.de, nilles@physik.tu-muenchen.de, 
theisen@mppmu.mpg.de}
\vskip1cm

\footnote{}{$^*$ Address after October 97: Physikalisches Institut, 
Unversit\"at Bonn, Nussallee 12, D-53115 Bonn.}

\footnote{}{$^{\diamond}$ Work supported in part by 
GIF--the German-Israeli Foundation for Scientific Research and
the European Commission TMR programmes ERBFMRX-CT96-0045,
CT96-0090.}
\abstract{We give a detailed discussion of the matching of the 
BPS states of heterotic, type I and type I$'$ theories in 
$d=9$ for general backgrounds. This allows us to explicitly identify
these (composite) brane states in the type I$'$ theory that lead to 
gauge symmetry enhancement at critical points in moduli space.
An example is the enhancement of $SO(16)\times SO(16)$ to 
$E_8\times E_8$. }

\vskip 2cm

\endpage

%
\noindent
{\sl Introduction}
\smallskip

In nine dimensions the heterotic theories and the orientifold projections
of the type II theories are believed to be related by a
chain of dualities. Already in ten dimensions it has been conjectured in 
\wvd\ and further substantiated in \pw\ that there is a strong-weak 
coupling 
duality between the heterotic $Spin(32)/\ZZ_2$ and the type I theory
\foot{Various aspects of heterotic - Type I duality have been 
verified in refs.\bk,\bfkov,\bd.}.
The latter, when viewed as an orientifold projection of the IIB theory,
lives on an orientifold nine-plane, and consistency requires
the presence of 16 pairs of D9 branes which give rise to an SO(32)  
gauge group \pl.
Upon compactification on a circle this becomes a nine-dimensional
duality. The type I theory on $S^1$ is \leigh\ 
T-dual to the orientifold
projection of the IIA theory known as I$'$  theory. The compact 
dimension of the I$'$ theory 
is topologically a segment and its two endpoints define two orientifold 
eight-planes. Consistency requires the presence of 16 mirror
pairs of D8 branes parallel to the orientifold planes.

In a background where the unbroken gauge group is SO(16)$\times$SO(16) 
one gets a closed chain of dualities.
For this background the two heterotic 
theories are related by $R\leftrightarrow 1/R$ duality \ginsp. 
Furthermore, one can view both, the nine-dimensional 
$E_8\times E_8$ heterotic theory and the type I$'$ theory as
compactifications of $M$ theory on a cylinder $S^1\times(S^2/\ZZ_2)$, 
where in the former
case the dilaton is related to the length of the cylinder, whereas
in the latter case it is related to its circumference. It has been 
conjectured in \hwone\ that the two 
theories are connected by 
a duality transformation which exchanges $S^1/\ZZ_2$ with $S^1$.
Attempts to describe the heterotic string as a matrix quantum mechanics 
of type I$'$ D particles \hetmatrix\ are based on this 
conjectured chain of dualities.

In this letter we present a detailed mapping of the 
type I$'$ D0 brane states to type I/heterotic
BPS states for generic backgrounds. The mapping is based on 
relating the BPS
mass formulas of the corresponding states over the whole moduli
space. In this way we can explicitly identify the objects 
in type I$'$ theory that are expected
to give rise to gauge symmetry enhancement at special 
points in the moduli
space. The generic picture that we find is that the 
heterotic/type I BPS states at
$n=1$ winding map to bound states of a single D0 brane 
sitting at one of the
orientifold planes with wrapped open and closed type I$'$ strings. 
In the SO(32) and other backgrounds where one observes a $U(1)\to SU(2)$
gauge symmetry enhancement at the self-dual heterotic/type I radius, one
can identify the $W^{\pm}$ gauge bosons with a 
D0 and anti-D0 brane sitting
at the orientifold plane which has no D8 branes. 
A D0 brane sitting at the opposite orientifold plane is 
the image of a spinor weight of the $Spin(32)/\ZZ_2$ lattice.

\medskip
%
%
\noindent
{\sl Review of heterotic-type I-type I$'$ duality}
\smallskip
Before matching the BPS states of the type I$'$ and the 
type I theories, or, via duality, of the type I$'$ and the heterotic
theory, we have to review some of the discussion of \pw.

The low energy effective action of the type I theory is
$$
\eqalign{
S_{I}&=\int\!d^{10}x \sqrt{-g}\,\coeff{1}{2\kappa_0^2}\,e^{-2\phi}
          \Bigl({\cal R}+4\,\partial_{M}\phi\,\partial^{M}\!\phi\Bigr)\cr
&\qquad   -T_9\int\!d^{10}x\sqrt{-{g}}\,e^{-\phi}\,
          [\,16+\coeff{(\pi\alpha^\prime)^2}{2}\,{\rm tr}_vF_{MN}F^{MN}]-
          16\,\mu_9\int\!A_{10}} 
\eqn\SIB
$$
The entire space is an orientifold plane which is charged under 
the ten-form field $A_{10}$. In addition there are also 16 D9 branes.
$F$ is the $SO(32)$ field strength, the trace
being in the vector representation  
and $\mu_p^2=2\kappa_0^2 T_p^2=2\pi(4\pi^2\ap)^{3-p}$, 
$\kappa_0=8\pi^{7/2}\ap^2 $ \pl.

The low energy effective action of the type I$'$ theory, 
on the other hand is, 
$$
\eqalign{
S_{I'}&=\int\!d^{10}x\sqrt{-g}\,\coeff{1}{2\kappa_0^2}\,e^{-2\phi}
                \left({\cal R}+4\,\partial_{M}\phi\,\partial^{M}\!\phi\right)-
                \coeff{1}{2}\int G_{10}\,^*\!G_{10}\cr
&\qquad  -\sum_i\Bigl\{T_8\int_{x^9=x_i^9}\!d^9x\sqrt{-{g}}\,e^{-\phi}\,
  [\,n_i+\coeff{(\pi\alpha^\prime)^2}{2}\,{\rm tr}_vF_{\mu\nu}F^{\mu\nu}]+
  n_i\mu_8\int_{x^9=x_i^9}\!A_9\Bigr\}}
\eqn\SIA
$$
The sum is over the D8 branes, with $n_i$ positioned at $x^9_i$; 
$\sum n_i=16$. $G_{10}$ is the field strength of the nine-form potential
to which the D8 branes couple; the gauge group depends on the 
positions of the D8 branes \pl. 

Polchinski and Witten have found a non-trivial background which solves the 
type I$'$ equations of motion:  
$$
\eqalign{
\kappa&=\kappa_0 e^{\phi'(y)}=z(y)^{-5/6}\,,\qquad
\Omega(y)=C z(y)^{-1/6}\,,\cr
z(y)&={3C\over\sqrt{2}}[B(y)\mu_8-\nu(y) y]\,,}
\eqn\background
$$
with $g_{MN}^{I'}=\Omega^2(y)\eta_{MN}$ and 
$y\in[0,2\pi]$ parameterizing
the segment $S^1/\ZZ_2$. The equation of motion for the 
10--form field strength $G_{10}(y)=\nu(y)dx^0\wedge...\wedge dx^9$ 
requires $\nu(y)$ to be piecewise constant with discontinuity 
$\Delta\nu=n_i\mu_8$ at a crossing of a group of $n_i$ D8 branes 
at $y=y_i$. In addition the type I$'$ projection implies the boundary 
conditions $\nu(2\pi)=-\nu(0)=8\mu_8$.
To get continuous metric and dilaton backgrounds, as required by
their equations of motion, $B(y)$ is also piecewise constant 
with discontinuity $\Delta B=n_i y_i$. 
$B(y)$ is fixed by $B\equiv B(0)$.
The type I$'$ space of classical vacua is thus parameterized
by the constants $B$ and $C$ together with the positions 
$y_i$ of the D8 branes. To have a meaningful background also requires
$z(y)\geq0$ everywhere. This implies (up to a physically irrelevant 
choice for the sign of $C$) that there is a minimum value $B_m$ for $B$
which depends on the configuration of D8 branes.

The type I$'$ theory is, by definition, the T-dual of the type I
theory. In a non-constant background, 
the relation between the 9--dimensional type I$'$ and type I metrics 
is not known a priori. Following \pw~ we set 
$g_{\mu\nu}^{I'}=\Omega^2(y)\gamma_{\mu\nu}=\Omega^2(y)
Q^2g_{\mu\nu}^{I}$, with $Q$ a constant factor to be determined.
Comparing the 9--dimensional gravitational actions gives
$$
2\pi\!R\,e^{-2\phi}=2Q^7\int_0^{2\pi}\!dy\Omega^8(y) e^{-2\phi'}=
2\kappa_0^2\,Q^7C^{25/3}\int_0^{2\pi}\!dy\left(
\coeff{z(y)}{C}\right)^{1/3}\ ,
\eqn\grav
$$
while comparison of the gauge actions results in
$$
R\,e^{-\phi}=\kappa_0\sqrt{\ap}\,C^5Q^5\ .
\eqn\gauge
$$
Note that since the integration on the right--hand--side of \grav~ is
restricted to $S^1/\ZZ_2$, there is an additional factor of 2 \wcy.

To establish the relation between $R$, $\phi$ and $B,\, C$ 
we need one additional equation, as we also have the unknown 
parameter $Q$. Still following ref.\pw, 
we will get this equation by comparing masses of BPS states in the two 
theories. 

Matching the masses of a type I Kaluza-Klein state of mass $1/R$ 
and a type I$'$ winding state of mass $m_{I'}$, taking into account the 
relation between the respective nine-dimensional metrics, i.e.
$1/R=Q m_{I'}$, one finds \pw
$$
{1\over R}={2Q\over 2\pi\ap} \int_0^{2\pi}\!\!\Omega^2(y)\, dy
          ={QC^{5/3}\over \pi\ap}\int_0^{2\pi}\!\!dy\left(\coeff{z(y)}{C}
            \right)^{-1/3}\ .
\eqn\third
$$
Solving \grav, \gauge~ and \third~ we find
$$
Q=C^{-5/6}{[\int_0^{2\pi}dy(\coeff{z(y)}{C})^{1/3}]^{1/4}\over
[\int_0^{2\pi}dy(\coeff{z(y)}{C})^{-1/3}]^{1/4}}
\eqn\Q
$$

$$
R=\pi\ap C^{-5/6}[\Int{0}{2\pi}dy(\coeff{z(y)}{C})^{-1/3}]^{-3/4}
[\Int{0}{2\pi}dy(\coeff{z(y)}{C})^{1/3}]^{-1/4}
\eqn\R
$$

$$
e^\phi=\coeff{1}{8}\pi^{-5/2}\ap^{-3/2}C^{-5/3}
[\Int{0}{2\pi}dy(\coeff{z(y)}{C})^{-1/3}]^{1/2}
[\Int{0}{2\pi}dy(\coeff{z(y)}{C})^{1/3}]^{-3/2}
\eqn\dilaton
$$
To complete the map between the vacua of type I$'$ 
and type I we need to 
establish the relation between a Wilson--line in type I and the positions 
of the D8 branes in type I$'$. In the absence
of a Wilson--line all D8 branes lie at $y=0$. We then introduce  
a Wilson line in the type I theory
with a single non--vanishing entry $A_i$; 
all type I gauge bosons whose 
roots have a non--vanishing $i$-th component will acquire a mass 
${A_i\over R}$ (we take here for simplicity $A_i>0$). 
This picture corresponds in the type I$'$ 
description to one in which all branes 
lie at $y=0$ except for one which lies at $y=y_i$; the mass of the gauge 
bosons with a non--vanishing entry in the 
$i$-th component of their root vector
comes, in this picture, from the stretching of open strings
between the branes at 0 and the one at $y_i$. Comparing the masses
one finds 
$$
A_i={RQC^{5/3}\over 2\pi\ap}\,\int_0^{y_i}\!dy
\left({z(y)\over C}\right)^{-1/3}
={1\over 2}{\int_0^{y_i}dy(\coeff{z(y)}{C})^{-1/3}\over
\int_{0}^{2\pi}dy(\coeff{z(y)}{C})^{-1/3}}
\eqn\wilson
$$
For $y_i=0,\,2\pi$ we have $A_i=0,\,1/2$, as expected.
\medskip
%
%
\noindent
{\sl Matching of the Spectra}
\smallskip
We will now show how heterotic winding modes, 
or, via heterotic-type I duality, the D-string winding modes of 
type I theory, map onto D0 brane states of the type I$'$ theory. 

The mass of a single D0 brane measured in type I units is
$$
M_{D0}(y)=Q\Omega(y){T_0\over 2}e^{-\phi'(y)}\ .
\eqn\dzeromass
$$
Q and $\Omega(y)$ appear as a result of the non-trivial relation of the
nine-dimensional metrics and the non-trivial world-line measure 
respectively. The factor 1/2 reflects the fact that a single D0 brane
has half the charge and therefore via the BPS condition half the tension
of a dynamical object which consists of a brane and its mirror 
image. 
Note that the mass of the D0 brane depends 
on its position in the compact dimension, which for a single D0 brane
has to be one of the two fixed points $y=0$ or $y=2\pi$.

A state in the $SO(32)$ heterotic string spectrum 
compactified on a circle $S^1$ of radius $R$ in the presence
of a Wilson line $A$ 
is characterized by its winding number $n$, its 
momentum quantum number $m$, a weight vector $p$ of 
the Spin(32)$/\ZZ_2$ lattice and oscillator excitations $N_{L,R}$. 
Its mass is ($a_R=0\,(\shalf)$ for R (NS) sector) 
(see e.g. \ginsp)
$$
\shalf\aph M^2=(N_L-1+\shalf p_L^2)+(N_R-a_R+\shalf p_R^2)
\eqn\hetmass
$$
with 
$$
\eqalign{
p_L&=\Bigl(p+A n,\sqrt{\coeff{\aph}{2}}
\bigl({m-A\cdot p-\shalf A^2 n\over R}+
{nR\over\aph}\bigr)\Bigr)\cr
p_R&=\sqrt{\coeff{\aph}{2}}\bigl(
{m-A\cdot p-\shalf A^2 n\over R}-
{nR\over\aph}\bigr)}
\eqn\lr
$$
Taking into account level matching, i.e. 
$N_L-1+\shalf p_L^2=N_R-a_R+\shalf p_R^2$ and restricting to
BPS states, i.e. $N_R=0\,(\shalf)$ for R (NS) sectors, 
one obtains
$$
M^2=\Bigl({m-A\cdot p-\shalf A^2 n\over R}
-{nR\over\aph}\Bigr)^2
\eqn\mass
$$
$$
N_L=1-nm-\shalf p^2\eqn\lema
$$
In these expressions $\aph$ is related to the heterotic  
string tension via $\aph=\coeff{1}{2\pi T_{\rm h}}$. 
It has been shown in \pw\ that a single type I closed D string 
has the worldsheet structure of the SO(32) heterotic string in 
the fermionic formulation. The tension of this string is 
$T_D=\coeff{1}{2}T_1e^{-\phi}=\coeff{1}{4\pi\ap}e^{-\phi}$.
Here $\ap$ is the type I scale and a factor of 1/2 appears again 
because the object we discuss is half a dynamical object.
We can then read the spectrum of a single type I D string by 
substituting in the previous formulas $T_D$ for $T_h$ or 
$2\ap e^{\phi}$ for $\aph$ .

To be specific, we will consider a Wilson line that breaks the 
gauge symmetry to $SO(16)\times U(8)\subset SO(32)$ so that at 
a generic value of $R$ the symmetry is 
$SO(16)\times U(8)\times U(1)\times U(1)$.
The choice of the Wilson line which accomplishes this is
$$
A=(\coeff{\epsilon}{2},\dots,\coeff{\epsilon}{2},0,\dots,0)
\eqn\wl
$$
with eight entries of each kind and $\epsilon\in[0,1]$. 
For $\epsilon=0$ we recover the 
$SO(32)$ string and for $\epsilon=1$ the gauge group is 
$SO(16)\times SO(16)$.
In the type I$'$ the 16 physical
D8 branes are split into two groups of eight, 
one positioned at $y=0$, the other at $y_0$ which follows from 
\wilson~ with $A_i=\coeff{\epsilon}{2}$.

D string winding modes are expected to map via T-duality to D0
brane states in type~I$'$, so
in a first step we are trying to determine those states of the 
D string which are mapped to a D0 brane
sitting at $y=0$ and $y=2\pi$ respectively. 
It is straightforward 
to show that 
for the choice of quantum numbers $N_L=0,\,n=m=1,\,p^2=0$ 
one has
$$
M^{I}={R\over2\ap}e^{-\phi}
-{1-\epsilon^2\over R}
=M^{I'}_{D0}(2\pi)
\eqn\matchf
$$
This in fact is the lightest type I state at $n=1$.
Note also that at the self-dual radius ${R_*}^2=2\ap(1-\epsilon^2)e^\phi$
this state together with the $n=m=-1$ one (which maps to an
anti-D0 brane sitting at $y=2\pi$)
become massless and provide the extra gauge bosons
for the enhancement $U(1)\to SU(2)$.
The $R=R_*$ locus in the type I/heterotic moduli space corresponds
to the $B=B_m$ locus on the type I$'$ side.
The type I state that matches $M^{I'}_{D0}(0)$ 
belongs to the spinor conjugacy class of Spin(32)$/\ZZ_2$.
Choosing the weight $p=(-\shalf,\dots,-\shalf)$ 
and quantum numbers $N_L=0,\,n=1,\,m=-1$
one indeed verifies
$$
M^I={(1-\epsilon)^2\over R}+{R\over 2\ap}e^{-\phi}=M^{I'}_{D0}(0)
\eqn\matchs
$$
This is again the lightest type I state at $n=1$ which belongs to the
spinor conjugacy class of Spin(32)$/\ZZ_2$, although for the specific
Wilson line that we chose it is not unique.
For the constant SO(16)$\times$SO(16) background, which corresponds
to $\epsilon=1$, we have $M^{I'}_{D0}(2\pi) = M^{I'}_{D0}(0)$ as expected.

The fact that the two conjugacy classes of Spin(32)$/\ZZ_2$ map
to D0 branes sitting at different ends of the type I$'$ compact 
dimension follows from T-duality; cf. also \pcj. 
The type I Wilson-line~
$A$ determines in type I$'$ the positions of the D8 branes via the 
dualization of 9-9 to 8-8 open strings (9-9 momenta in the compact 
direction
become 8-8 windings). The position of a D0 brane in type I$'$ is 
determined relative to the position of the D8 branes via the
dualization of 1-9 to 0-8 open strings. To be concrete
let us consider the case with unbroken SO(32), when all the 
type I$'$ D8 branes
sit at $y=0$. In type~ I the 1-9 strings can feel the presence
of a gauge holonomy around a closed D string. The possible gauge 
holonomies belong to the $\ZZ_2$ group $\{1,-1\}$ \pw.
The massless excitations of the 1-9 strings live in the
world-volume of the D string and are fermionic degrees of
freedom with respect to the SO(1,1) Lorentz group of this
world-volume theory \pw. The transportation of a massless 1-9 state around 
the closed D string detects the presence or absence of a non-trivial
holonomy and induces a $-1$ or $+1$ phase respectively.
This means that in the presence of a non-trivial holonomy the massless
1-9 excitations give rise to anti-periodic fermions on the D string 
world-volume, while the trivial holonomy results in periodic ones.
When the
D string is wrapped once around the compact dimension, and in the 
absence of a gauge holonomy, the momentum mode of the 1-9 strings
along this direction is integer moded (in quanta of $1/R$),
while in the presence of the non-trivial $-1$ holonomy it is
half-integer moded. This maps via T-duality to integer and
half-integer windings of the 0-8 strings in the type I$'$ theory
respectively, which in turn implies that the wrapped D string with
a trivial holonomy maps to a D particle 
at $y=0$, that is on top of the D8 branes, 
while the non-trivial holonomy puts
the D particle at $y=2\pi$ making the 0-8 windings half-integral.
Given then the fact that the trivial holonomy corresponds
to the periodic sector of the D string current algebra, which
provides the spinor conjugacy class of Spin(32)$/\ZZ_2$, one expects
that this conjugacy class be represented in type I$'$ theory by a 
D0 brane sitting at $y=0$. Accordingly the adjoint conjugacy class
which arises in the anti-periodic sector of the D string current 
algebra should be represented by a D0 brane sitting at $y=2\pi$.
The same conclusions about the position of a single D0 brane and 
its expected relation to weights in the spinor or adjoint conjugacy class
of Spin(32)$/\ZZ_2$ arise in the more complicated case when a type I 
Wilson line is present moving some or all of the D8 branes away from 
$y=0$.

Beyond the two heterotic states that match the mass of a D0 brane 
placed at $y=2\pi$ and at $y=0$ respectively,
the heterotic string has a whole host of additional 
BPS states at $n=1$. Let us first however recollect what happens at $n=0$.
For $n=0$ the level-matching condition \lema~ implies that
the BPS spectrum is exhausted by the two possibilities
$p^2=0$, $N_L=1$ and $p^2=2$, $N_L=0$  with  arbitrary values of the
Kaluza-Klein number $m$. This is nothing but the tower of states resulting 
from the Kaluza-Klein reduction of the ten-dimensional
supergravity and super-Yang-Mills multiplets to nine dimensions.
The mass of these states is always proportional to $1\over R$.
At the lowest level ($m=0$) it is zero for the gravity and 
the unbroken gauge multiplets, and 
${1\over R}A\cdot p$ for the SO(32) gauge bosons with $A\cdot p\neq 0$.
All of these states are mapped in the type I$'$ theory to winding
modes of closed and open fundamental strings. The heterotic Kaluza-Klein 
momentum maps to type I$'$ winding number. 
The class of heterotic states with  
$p^2=0$, $N_L=1$ map on the I$'$ side to the lowest level of stretched closed
strings if they belong to the supergravity multiplet,
or to the lowest level of stretched open strings with endpoints
on the same D8 brane if they are in the Cartan subalgebra.
The second class of heterotic BPS states ($p^2=2$, $N_L=0$) 
correspond to non-vanishing roots of the SO(32) lattice and are mapped
in type I$'$ 
to the lowest level of wound open strings with ends on different
D8 branes.

At $n=1$ the picture is more complicated. Level-matching now permits 
arbitrarily large values of both $N_L$ and $p^2$, since their contribution
can be canceled by negative values of $m$. Starting with 
$p^2=0$, $N_L=0$ we get $m=1$ and this is the largest possible value of $m$ 
at $n=1$, corresponding to the lightest state at this winding, as can be seen
from \mass. We
identified this state with a D0 brane sitting at $y=2\pi$ in the type I$'$ 
theory. Further inspection of equations \mass\ and \lema\ 
shows that increasing
$N_L$ by one unit has to be compensated for level-matching by a one-unit 
decrease in $m$. This results in an 
increase\foot{Note that we work with radii larger than the self-dual 
radius for the type I$'$ background to be meaningful, i.e. to have 
$z(y)\geq 0$.}
in the mass of the state by $1\over R$. Similarly increasing $p^2$ by two has
to be compensated by decreasing $m$ by one and the 
corresponding increase 
in the mass of the state is now equal to $\coeff{1+A\cdot\Delta p}{R}$,
where $\Delta p$  is the difference between the new and the original
weight. 
It is then evident that all mass increases above the lightest state
are proportional to $1\over R$, a scale which maps on the
type I$'$ side to the scale of fundamental string windings.
These observations suggest the following type I$'$ picture:
Starting with the lightest object at $n=1$, which is the
D0 brane at $y=2\pi$, one could envisage constructing 
the states which match the
heterotic BPS states with $N_L>0$, as bound states at threshold 
of the given D0 brane
with closed and open (ends on the same D8 brane) strings 
at positive winding\foot{Negative
winding strings would then bind with the anti-D0 brane to provide the
corresponding heterotic states at $n=-1$.}. These stretched strings
have in their ground state a mass which equals their winding number times 
$1\over R$ in heterotic units, as suggested by equation \third,
and therefore reproduce exactly the increase in the heterotic BPS mass
brought about by non-vanishing values of $N_L$.  
The heterotic states with $p^2>0$ could then be matched by bound
states at threshold of the D0 brane with wound 
open strings whose ends are on different D8 branes. 
Here and below we assume the existence of these threshold bound states.
For example the heterotic BPS states with $p^2=2$, $N_L=0$ have 
$m=1$ and exceed the
mass of the D0 brane at $y=2\pi$ by an amount 
by $\coeff{1+A\cdot p}{R}$ (in heterotic units).
This excess of mass is 
exactly the stretching energy of an open string which winds once
around the compact dimension
(i.e. from 0 to $2\pi $ and back),  
and has its ends fixed on the pair of D8 branes
which correspond to the considered root 
$p=\pm(e_i\pm e_j),\, i,j=1,\dots, 16$. 
There is however a subtlety here if we want to adopt the interpretation
that the non-trivial gauge quantum numbers of these type I$'$ states
are due to the Chan-Paton indices of the open string part in the bound
state: one can never represent a
spinor weight of Spin(32)$/\ZZ_2$ with open strings stretched between
D8 branes. This is the T-dual of the observation that one does not get 
the spinor of Spin(32)$/\ZZ_2$ in the open string sector of the type I
theory. What this suggests is that it is not possible to start with the 
D0 brane
at $y=2\pi$ and construct the whole set of heterotic BPS states at $n=1$
by binding this D0 brane with stretched open and closed strings.
This however is just as well since we saw that a D0 brane placed at
$y=0$ has exactly the right mass to represent the lightest heterotic states
in the spinor of Spin(32)$/\ZZ_2$. 
Given then the fact that the difference of two spinorial weights is
always a weight in the 0 conjugacy class, one expects that 
bound states of the D0 brane at $y=0$ with stretched open strings will
represent the whole set of heterotic $n=1$ states
in the spinor of Spin(32)$/\ZZ_2$.
The picture just described reproduces in the I$'$ theory 
the heterotic $n=1$ BPS mass formula.
At the same time it accounts for the
gauge quantum numbers of these states in I$'$.

The SU(2) symmetry 
enhancement which occurs in the heterotic theory 
at the 
self-dual radius for generic backgrounds comes about by means of 
the two states $n=m=1$ and $n=m=-1$ at $p=N_L=0$.
We have argued that these states map to a D0 and an anti-D0 brane 
respectively placed at $y=2\pi$. It is noteworthy that when 
attempting 
to understand the SU(2) enhancement in the type I$'$ both
a D0 and an anti-D0 brane seem to be essential ingredients, 
whereas in the proposed matrix description of the strong coupling limit of 
the IIA theory in the infinite momentum frame only D0 branes 
are required \bfss. For the special choice of background for which
the unbroken gauge group is SO(16)$\times$SO(16)
the self-dual radius is zero and the symmetry enhancement which 
occurs there is more interesting: a large number of states become 
massless at every value of $n$, thus enhancing the unbroken gauge group
to $E_8\times E_8$ and at the same time unfolding a new large dimension
\foot{An alternative approach to recovering 
the $E_8\times E_8$ symmetry in the presence of a 
$SO(16)\times SO(16)$ background can be found in ref.\bd.}.
At $n=1$ the states that become massless are in the (128,1)+(1,128) 
of SO(16)$\times$SO(16) \ks, and the state with $n=m=1$ at
$p=N_L=0$ is one of them. It is interesting that for this choice of 
background we can consistently consider only positive values
of $n$ (this implies that we restrict ourselves to positive
momenta along the new dimension which emerges). 
For generic backgrounds on the other hand, 
one needs states at $n=\pm1$ to get the 
$W^\pm$ bosons for the gauge symmetry enhancement 
$U(1)\to SU(2)$.

Heterotic BPS states at windings $n>1$ map to type I$'$ 
bound states of $n$ D0 branes at a fixed point with wrapped
closed and open fundamental strings. An interesting observation is that 
the spectrum of the heterotic theory does not contain BPS states at $n>1$
whose mass can match bound states at threshold of $n$ D0 branes only,
i.e. without wrapped fundamental strings.

\vskip.3cm

We thank J. Polchinski for a useful email exchange and the Theory Division at 
CERN for hospitality during the initial and the final stages of this 
work.

\refout
\end